\documentclass[11pt]{article}

\usepackage{epsfig,subfigure}
\usepackage{amssymb, amsmath}
\usepackage[usenames,dvipsnames]{color}

\def\QED{\mbox{\rule[0pt]{1.5ex}{1.5ex}}}
\def\proof{\noindent\hspace{2em}{\it Proof: }}

\usepackage{graphicx}

\usepackage{amssymb}
\usepackage{amsmath,amsfonts,amssymb}
\usepackage{verbatim}
\usepackage{stfloats}
\usepackage[bookmarks=false]{}

\newtheorem{theorem}{Theorem}

\newtheorem{lemma}{Lemma}

\begin{document}
%
%

\title{Degrees of Freedom of MIMO $X$ Networks: \\  Spatial Scale Invariance, One-Sided Decomposability and Linear Feasibility\thanks{This work was presented in part at ISIT 2012 \cite{mimox_conf}. The work was supported  by ONR grant N00014-12-10067,  by NSF CCF-1161418 and by a gift from Broadcom.}}
\author{\normalsize  Hua Sun, Chunhua Geng, Tiangao Gou and Syed A. Jafar\\
        {\small Department of Electrical Engineering and Computer Science}\\
        {\small University of California, Irvine, Irvine, CA 92697}\\
      {\small \it E-mail~:~\{huas2, chunhug, tgou, syed\}@uci.edu}\\
       }
       

\date{}

\maketitle

\begin{abstract}

We show that an $M\times N$ user MIMO $X$ network with $A$ antennas at each node has $A\left(\frac{MN}{M+N-1}\right)$ degrees of freedom (DoF), thus resolving in this case a discrepancy between the spatial scale invariance conjecture (scaling the number of antennas at each node by a constant factor will scale the total DoF by the same factor) and a decomposability property of over-constrained wireless networks. While the best previously-known  general DoF outer bound is consistent with the spatial invariance conjecture, the best previously-known   general DoF inner bound, inspired by the $K$ user MIMO interference channel, was based on the decomposition of every transmitter and receiver into multiple single antenna nodes, transforming the network into an $AM\times AN$ user SISO $X$ network. While such a decomposition is DoF optimal for the $K$ user MIMO interference channel, a gap remained between the best inner and outer bound for the MIMO $X$ channel. Here we close this gap with the new insight that the MIMO $X$ network is only one-sided decomposable, i.e.,  either all the transmitters or all the receivers (but not both) can be decomposed by splitting multiple antenna nodes into multiple single antenna nodes without loss of DoF. The result is extended to SIMO and MISO $X$ networks as well and in each case the DoF results satisfy the spatial scale invariance property. In addition, the feasibility of linear interference alignment is investigated based only on spatial beamforming without symbol extensions. Similar to MIMO interference networks, we show that when the problem is improper, it is infeasible.
\end{abstract}

\newpage

\section{Introduction}
The use of multiple antennas, known as multiple input multiple output (MIMO) technology, and the consolidation of interference, known as interference alignment (IA), are two of the promising advances of the last two decades that seek to alleviate the spectrum shortage for wireless communication networks by making available additional spatial degrees of freedom (DoF). Taken individually, the understanding of MIMO is by now quite mature, and rapid advances have recently been made in understanding the essential principles of IA through DoF studies of a variety of network settings. Taken together, however, the understanding of MIMO in conjunction with IA -- the understanding of the spatial dimension per se -- is limited by a number of unresolved fundamental issues. In terms of systematic insights, a number of properties have been identified in \cite{Jafar_ITA2012} that are true for all known DoF results, and conjectured to be true in general (i.e., for almost all channel realizations), but for which a general proof (or counter-example) is not yet known. These observations include the duality property (reciprocal networks have the same DoF), the diversity property (time-varying channels have the same DoF as constant channels), the linearity property (linear beamforming schemes over time-varying channels are sufficient to achieve the information theoretic DoF), and especially relevant to this work, the properties of spatial scale invariance and decomposability. Remarkably, these two properties, which hold for all DoF results known previously, contradict each other for MIMO $X$ networks. Resolving this  curious discrepancy is an open problem that is highlighted in  \cite{Jafar_tutorial} (Page 81, Sec. 5.4). The main motivation of this work is to solve this open problem.  

\subsection{Spatial Scale Invariance and Decomposability}

\subsubsection{Spatial Scale Invariance}
It is well understood that the DoF of wireless networks are scale-invariant with respect to time and frequency dimensions. Wang et al. have recently conjectured in \cite{Wang_Gou_Jafar_3mimoic} that the spatial dimension is similarly scale invariant:

``\textit{With perfect global channel knowledge and generic channels, if the
number of antennas at each node in a wireless network is scaled by a common constant factor, then the DoF of the network (for almost all channel realizations) scale by the same factor.}"

The spatial scale invariance conjecture is consistent with \emph{all} known DoF results across a wide variety of networks, which includes interference networks, $X$ networks, cellular networks, and even multi-hop networks \cite{Jafar_tutorial}. In particular, we note that for the $2\times 2$ user MIMO $X$ channel with $A$ antennas at each node, the DoF value is known to be $\frac{4A}{3}$ \cite{Jafar_Shamai_X, Cadambe_Jafar_Wang}, which scales with $A$ and is therefore, consistent with the spatial scale invariance conjecture. Even for the $M\times N$ user MIMO $X$ network (i.e., an $X$ network with $M$ transmitters and $N$ receivers) with $A$ antennas at each node, if $\min(M,N)\leq 2$, the DoF value is easily seen to be $A\left(\frac{MN}{M+N-1}\right)$, again spatial scale invariant \cite{Cadambe_Jafar_X}. However, if $\min(M,N)> 2$, the DoF remain unknown. Interestingly, the best known DoF outer bound for this setting is consistent with the spatial invariance conjecture \cite{Cadambe_Jafar_X}.



\subsubsection{Decomposability}
We use the term ``decomposition" to refer to independent processing at each antenna, essentially splitting a multiple antenna node into multiple independent single antenna nodes.  It was first used to simplify the proof of achievability in the $K$ user symmetric (equal number of antennas at all nodes) MIMO interference channel \cite{Cadambe_Jafar_int} where the DoF result obtained for the SISO setting was immediately extended to the symmetric MIMO setting by decomposing the $K$ user MIMO interference network with $A$ antennas at each node, into an $AK$ user SISO interference network, where the asymptotic CJ alignment scheme \cite{Jafar_tutorial, Cadambe_Jafar_int} can be applied to show that $AK/2$ DoF are achievable, without joint processing among co-located antennas at any node. Since $AK/2$ is also the DoF outer bound for the $K$ user symmetric MIMO interference network, it is evident that the network is decomposable, i.e., no loss of DoF results from decomposing all transmitters and receivers. The decomposability property is also known to be true for $K$ user MIMO interference networks with $A_t$ antennas at each transmitter and $A_r$ antennas at each receiver, giving us the optimal (information theoretic) DoF value of $\frac{A_tA_r}{A_t+A_r}$ per user, provided that $K\geq\frac{A_t+A_r}{\mbox{gcd}(A_t,A_r)}$ \cite{Gou_Jafar_mimoint, Ghasemi_Motahari_Khandani_mimoic}. Further study of the $K$ user MIMO interference channel by Wang et al. leads to the conjecture that decomposability holds in all \emph{over-constrained} (also known as improper \cite{IC_feasibility_TSP}) settings, i.e., where the information theoretic DoF value per user is higher than $\frac{A_t+A_r}{K+1}$. Based on previously existing DoF results, a general pattern   summarized in \cite{Jafar_tutorial}  states that:

``\emph{The DoF benefits of collocated antennas disappear with increasing number of alignment constraints}".

Evidently, this is because for over-constrained networks the multiplicity of alignment constraints invariably requires the use of the CJ scheme \cite{Jafar_tutorial, Cadambe_Jafar_X, Cadambe_Jafar_int}, which does not require  joint processing across multiple antennas, instead  breaking them into separate
nodes. The CJ scheme is inherently a decomposition based scheme because of its reliance on commutativity of channel matrices, a property satisfied by the diagonal channels obtained by time/frequency symbol extensions of SISO channels, but not by time/frequency extensions of MIMO channels (which would only produce non-commuting block-diagonal channels).

The previously best known inner bound for  $M\times N$ user MIMO $X$ network with $A$ antennas at each node, and $\min(M,N)>2$,  is also based on the decomposition argument and application of the asymptotic CJ alignment scheme \cite{Cadambe_Jafar_X}. By decomposing every transmitter and receiver in an $M\times N$ user  MIMO $X$ network with $A$ antennas at each node, we obtain an $AM\times AN$ user  SISO $X$ network, and therefore the corresponding DoF value, $A\left(\frac{MN}{M+N-\frac{1}{A}}\right)$ is achievable \cite{Cadambe_Jafar_X}.

\subsection{Summary of Contribution}
The main goal of this work is to resolve, in the context of MIMO $X$ networks, the apparent discrepancy between the spatial invariance conjecture, as represented by the best available DoF outer bound, and the decomposability property, as represented by the best available DoF inner bound. As mentioned above, for $M\times N$ user MIMO $X$ network with $A$ antennas at each node, and with $\min(M,N)>2$, there remains a gap between the best DoF outer bound value, $A\left(\frac{MN}{M+N-1}\right)$, and the best DoF inner bound value, $A\left(\frac{MN}{M+N-\frac{1}{A}}\right)$. This gap represents an opportunity to refine our  understanding of the spatial invariance and decomposability properties. While the gap may seem small for large values of $A$, note that because DoF is a very coarse metric, even a small gap between DoF bounds corresponds to unbounded gaps in the corresponding capacity bounds. To summarize the motivation for this work, MIMO $X$ networks represent an important class of wireless networks, a precise DoF characterization  is highly desirable, it would close the open problem highlighted in \cite{Jafar_tutorial}, and improve our understanding of the fundamental structure of signal dimensions.

The main contribution of this work is the precise DoF characterization for $M\times N$ user MIMO $X$ networks with $A$ antennas at each node (and all SIMO and MISO $X$ network settings). This involves both new insights as well as new technical challenges. In terms of new insights, we settle the spatial invariance conjecture for MIMO $X$ networks with $A$ antennas at each node, i.e., we show that the DoF outer bound is tight, also closing the heretofore open DoF problem for these networks. The discrepancy with the previous inner bound is resolved by  improving our understanding of the decomposability property. We find that, unlike MIMO interference networks which demonstrate a two-sided decomposability, i.e., both the transmitters and receivers can be decomposed into single antenna nodes, MIMO $X$ networks are only one-sided decomposable, i.e., either the transmitters or the receivers (but not both simultaneously) can be decomposed into single antenna nodes without loss of DoF.
Interestingly, this is not because of the alignment constraints. Indeed the alignment still takes place very much like a SISO setting, based entirely on the CJ scheme. Instead, this is because of the separability of desired and interference signals. As it turns out, joint processing at one end, e.g., at the receivers in a SIMO X network, allows a larger space within which the desired signals can be resolved more efficiently from the interference. The use of the CJ scheme for achievability is significant because the same scheme often translates into the rational dimensions framework to establish corresponding DoF results in static settings (see, e.g., \cite{Jafar_tutorial, Motahari_Gharan_Khandani, Ramamoorthy_Wang_Yin}). Indeed, the DoF results of this paper have been recently extended to constant settings by  Zamanighomi and Wang in \cite{Wang_joint}. One-sided decomposability features prominently in \cite{Wang_joint} as well.

While the new insights are the  key ingredient to closing this open problem, there are non-trivial technical challenges involved as well. In particular, the mathematical proof of the resolvability of desired signals from interference with joint processing across the non-decomposed receivers (the reciprocal setting follows by duality), poses new challenges. This requires proving the full rank property of a matrix (signal space matrix) whose columns represent the received signal vectors and whose rows represent the receive antennas and channel uses. What complicates matters is that this matrix contains dependencies across both rows and columns. The dependencies across rows arise because of the multiple receive antennas that receive different linear combinations of the same set of symbols over each channel use. The dependencies across columns arise due to the $X$ setting, because each channel coefficient is involved with both desired and interfering signals. When taken individually, the dependencies across rows have been addressed in the MIMO interference network setting  by Gou and Jafar in \cite{Gou_Jafar_mimoint} and the dependencies across columns have been addressed in the $X$ network setting by Cadambe and  Jafar in \cite{Cadambe_Jafar_X}. However, as it turns out, dealing with both kinds of dependencies simultaneously is especially challenging. Our proof relies on a slightly modified CJ scheme, and uses  mathematical induction to construct the overall signal space matrix in a stepwise manner by appending blocks of rows and columns while at each stage proving that this does not introduce rank deficiencies.

Finally, as a relatively minor addendum, we visit the issue of linear feasibility that has previously attracted much research interest for MIMO interference networks, and study it here in the context of MIMO $X$ networks. Linear feasibility refers to the achievability of interference alignment in a wireless network based only on spatial beamforming, i.e., without symbol extensions. Starting with a general formulation of the feasibility of linear interference alignment by Gomadam et al. in the context of interference networks \cite{Gomadam_Cadambe_Jafar_dist}, the feasibility question was explored by Cenk et al. in \cite{IC_feasibility_TSP} in terms of the solvability of a set of multivariate polynomial equations, leading to the categorization of an IA problem as improper or proper based on whether or not the number of equations exceeds the number of involved variables, and the conjecture that proper systems (combined with information theoretic bounds) are likely to be feasible and improper systems are likely to be infeasible.  This conjecture is recently settled completely in one direction and partially in the other direction in several recent works \cite{Wang_Gou_Jafar_3mimoic, IC_feasibility_Tse, IC_feasibility}. In particular, it has been shown that in interference channel setting improper systems are infeasible. Here, we extend this result to the MIMO $X$ setting. Following the approach of \cite{IC_feasibility_Tse, IC_feasibility}, we establish that in arbitrary (not limited to symmetric) MIMO $X$ networks, the improperness of the underlying polynomial system implies infeasibility of linear IA as well.

\section{System Model}
An $M\times N$ user MIMO $X$ network is a single-hop communication network with $M$ transmitters and $N$ receivers, where transmitter $i$ has message $W^{[ji]}$ for receiver $j$, for each $i \in \{1,2,\ldots, M\}, j \in \{1,2, \ldots, N\}$. Transmitter $i$ has $A_i$ antennas and receiver $j$ has $B_j$ antennas. The $M\times N$ user MIMO $X$ network is described by input-output relationship
$$\mathbf{Y}^{[j]}(\kappa) = \displaystyle\sum_{i \in \{1,2,\ldots,M\}} \mathbf{H}^{[ji]}(\kappa) \mathbf{X}^{[i]}(\kappa) + \mathbf{Z}^{[j]}(\kappa),~~~{ j\in\{1,2,\ldots, N\}}$$
where $\kappa$ represents the channel use index, $\mathbf{X}^{[i]}(\kappa)$ is the $A_i \times 1$ input signal vector of the $i^{th}$ transmitter, $\mathbf{Y}^{[j]}(\kappa)$ is the $B_j \times 1$ output signal vector of the $j^{th}$ receiver and $\mathbf{Z}^{[j]}(\kappa)$ represents the $B_j \times 1$ additive white Gaussian noise (AWGN) vector at the $j^{th}$ receiver. The average transmit power at each transmitter is bounded by $\rho$ (referred to as the Signal-to-Noise Ratio) and the i.i.d. noise variance at all receivers is assumed to be equal to unity. $\mathbf{H}^{[ji]}(\kappa)$ represents the $B_j \times A_i$ channel matrix between transmitter $i$ and receiver $j$ at channel index $\kappa$. We assume that all channel coefficient values are time-varying, i.i.d., drawn from a continuous distribution and the absolute value of all the channel coefficients is bounded between a non-zero minimum value and a finite maximum value. Perfect knowledge of all channel coefficients is available to all transmitters and receivers. Let $R_{ji}(\rho) = \frac{\log|W^{[ji]}(\rho)|}{\kappa_0}$ denote the rate of the codeword encoding the message $W^{[ji]}$, where $|W^{[ji]}(\rho)|$ is the size of the message set and $\kappa_0$ is the length of the codeword. The rate $R_{ji}(\rho)$ is said to be achievable if for message $W^{[ji]}$, the probability of error can be made arbitrarily small with appropriately large $\kappa_0$. The closure of all achievable rate tuples is known as the capacity region. The DoF for message $W^{[ji]}$ is defined as $d_{ji}=\lim_{\rho \to \infty} R_{ji}(\rho)/\log(\rho)$, which can be interpreted as the number of independent signaling dimensions available for $W^{[ji]}$. Analogous to the capacity region, the DoF region, $\mathcal{D}$, is the closure of the set of all achievable DoF tuples. The sum-DoF value is defined as $\max_{\mathcal{D}}\sum_{1\leq i\leq M, 1\leq j\leq N} d_{ji}$. The symmetric DoF is the highest value $d$, such that the DoF allocation $(d,d,\cdots, d)$, is inside the DoF region. 


\section{Results}
In this section we present the statements of the main results along with some expository discussion. The proofs are relegated to the next section.
\subsection{Spatial Scale Invariance and Decomposability}
The main result is presented in the following theorem.
\begin{theorem}
The $M\times N$ user MIMO $X$ network with $A$ antennas at each node has $A\left(\frac{MN}{M+N-1}\right)$ total DoF almost surely.
\end{theorem}

\begin{figure}[!h]
\centering
\includegraphics[width=5.1in]{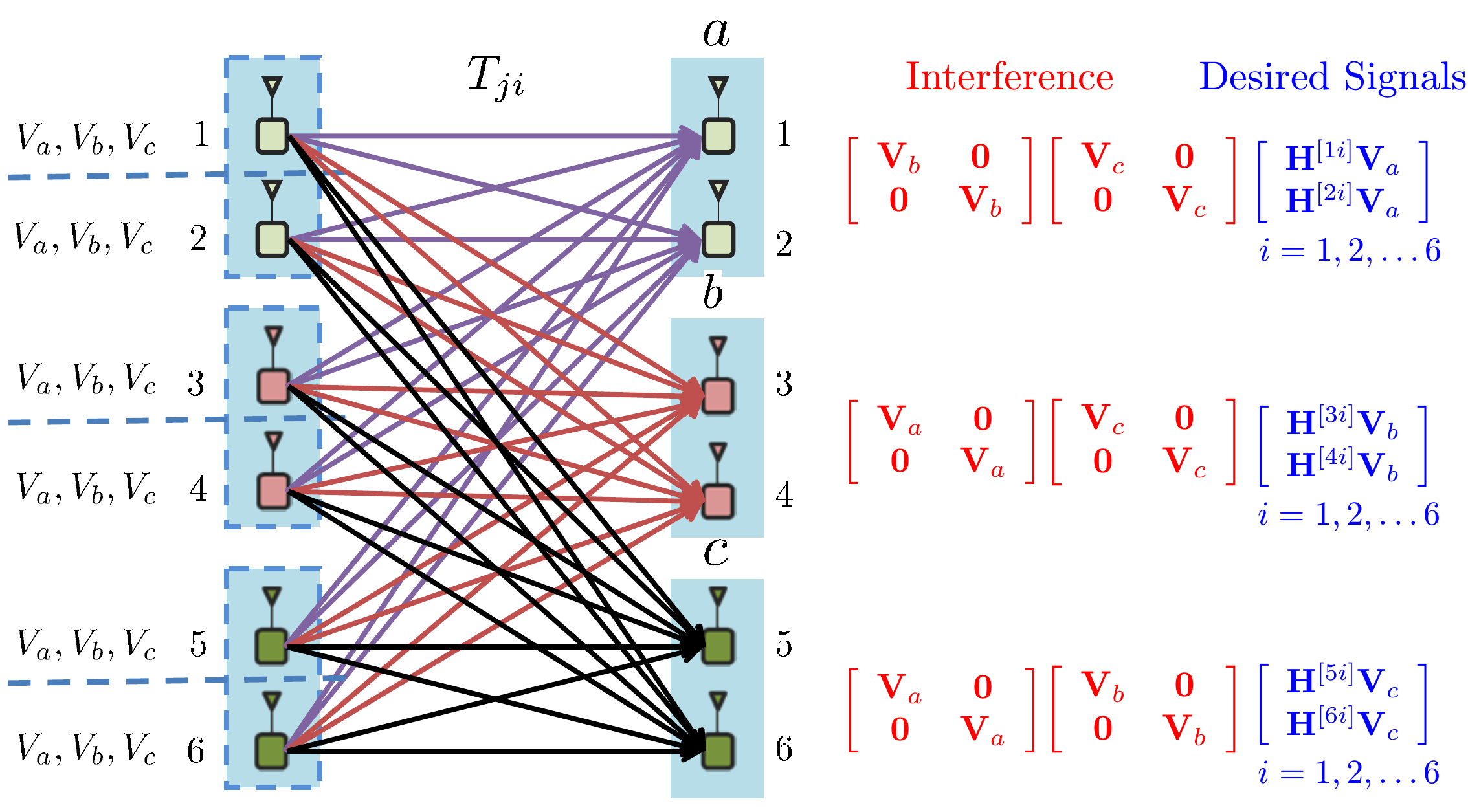}
\caption{IA after one-sided decomposition on the $3 \times 3$ MIMO $X$ channel}
\label{fig1}
\end{figure}

While a detailed proof appears in Section \ref{sec:inner}, let us convey the essence of the achievable scheme through a simple example  presented in a manner consistent with  the tutorial \cite{Jafar_tutorial}. Consider the $3 \times 3$ MIMO $X$ network with $2$ antennas at each node, i.e., $M=N=3,A=2$, as shown in Figure \ref{fig1}.
We split all the transmit antennas and view them as $6$ independent transmitters. Each virtual transmitter selects the same beamforming matrices (thereby same signal space) $\mathbf{V}_{a},\mathbf{V}_{b},\mathbf{V}_{c}$, for receiver $a,b,c$, respectively, over $n$ symbol extensions. The signal space seen by each receiver has $2n$ dimensions. Consider the symbols desired by receiver $a$, which constitute interference at receivers $b, c$. These symbols are transmitted by each transmitter along the signal space designated as $\mathbf{V}_a$. Note that because each receiver has $2$ antennas while each (decomposed) transmitter has only $1$ antenna, the symbols sent from any two transmitters cannot align with each other at any receiver. In other words, one-to-one alignments are not possible. Therefore, in order to consolidate the interference caused by $\mathbf{V}_{a}$ at receivers $b,c$ as much as possible, we turn to a many-to-many alignment scheme. Specifically, the $\mathbf{V}_a$ spaces from  transmitters $1,2$ occupy $2|\mathbf{V}_{a}|$ dimensional interference space at each undesired receiver (receivers $b$ and $c$) and all the remaining undesired $\mathbf{V}_a$ space signals sent from transmitters $3$ to $6$ are incorporated into these $2|\mathbf{V}_{a}|$ dimensions.  As shown in Figure \ref{fig1}, let us set the interference space occupied by signals sent along $\mathbf{V}_{a}$, as seen by receivers $b,c$, to $\mathbf{V}_{a} \times \mathbf{V}_{a}$, i.e., the column span of the matrix
\begin{equation}
\left[\begin{array}{cc}\mathbf{V}_{a} & \mathbf{0}\\
\mathbf{0} & \mathbf{V}_{a} \end{array}\right]
\end{equation}
in which all interference will be aligned, i.e.,
\begin{equation}
\text{span}\left[\begin{array}{c}\mathbf{T}_{3i}\mathbf{V}_{a} \\
\mathbf{T}_{4i}\mathbf{V}_{a} \end{array}\right] \subset
\text{span}\left[\begin{array}{cc}\mathbf{V}_{a} & \mathbf{0}\\
\mathbf{0} & \mathbf{V}_{a} \end{array}\right]
,\forall i=1, 2, \ldots, 6
\end{equation}
and
\begin{equation}
\text{span}\left[\begin{array}{c}\mathbf{T}_{5i}\mathbf{V}_{a} \\
\mathbf{T}_{6i}\mathbf{V}_{a} \end{array}\right] \subset
\text{span}\left[\begin{array}{cc}\mathbf{V}_{a} & \mathbf{0}\\
\mathbf{0} & \mathbf{V}_{a} \end{array}\right]
,\forall i=1, 2, \ldots, 6
\end{equation}
where $\mathbf{T}_{ji}$ denote the interference-carrying matrices (same as $\mathbf{H}^{[ji]}$  in the figure). All of these alignment conditions can be satisfied with the CJ alignment scheme \cite{Jafar_tutorial}\footnote{The notation ${\bf V}_a\approx{\bf T}_{ji}{\bf V}_a$ means that $\frac{\mbox{dim}(\mbox{span}({\bf V}_a)\cap\mbox{span}({\bf T}_{ji}{\bf V}_a))}{\mbox{dim}(\mbox{span}({\bf V}_a)\cup\mbox{span}({\bf T}_{ji}{\bf V}_a))}$ asymptotically approaches $1$.}
\begin{equation}
\mathbf{V}_{a} \approx \mathbf{T}_{ji}\mathbf{V}_{a}, \forall i=1, 2, \ldots, 6,j=3,4,5,6.\label{eq:Va}
\end{equation}
Similarly, the three messages for receiver $b$ are sent along the same signal space $\mathbf{V}_{b}$ by each transmitter, and align into the same space $\mathbf{V}_{b} \times \mathbf{V}_{b}$ at receivers $a,c$ where they constitute interference. Lastly, $\mathbf{V}_{c} \times \mathbf{V}_{c}$ spans the total interference space due to the messages intended for receiver $c$ , as seen by receivers $a,b$. The size of the signal spaces are chosen to be equal, i.e.,
$|\mathbf{V}_{a}|=|\mathbf{V}_{b}|=|\mathbf{V}_{c}|=|\mathbf{V}|$, and  $|\mathbf{V} \times \mathbf{V}|=2|\mathbf{V}|$. We can easily see that at each receiver, desired signals from all 6 transmitters occupy $6|\mathbf{V}|$ dimensions, and interference (namely signals intended for the other two receivers) occupies $4|\mathbf{V}|$ dimensions. After aligning the interference, we need to guarantee the linear independence of desired signals from interference. This is proved in Section \ref{sec:inner}.


For the desired and interference spaces to be linearly independent, we need the size of the total signal space, $2n$,  to be big enough to accommodate both. This is accomplished by setting $10|\mathbf{V}|=2n$. The total accessible DoF for the network equal $\frac{3 \times 6|\mathbf{V}|}{n} = \frac{18}{5} = A \left( \frac{MN}{M+N-1}\right)$, as desired.\\

The statement of Theorem $1$ can be further generalized to  SIMO and MISO settings, as in Theorem $2$.
\begin{theorem}\label{thm:2}
{\it The $M\times N$ user SIMO $X$ network with a single antenna at each transmitter and $R$ antennas at each receiver, as well as its reciprocal channel, the $N\times M$ MISO $X$ network, almost surely has a total of 
\begin{eqnarray*}
\mbox{DoF} = \min\left(M, \frac{MNR}{M+NR-R}\right).
\end{eqnarray*}
 In addition, the DoF in both cases satisfy the spatial scale invariance property.}
\end{theorem}
Note that Theorem $1$ is a special case of Theorem $2$ when $R=1$ and the scaling factor is specified by $A$.

\subsection{Linear Feasibility}\label{linear_feasibility}
When considering linear IA without symbol extension, we assume transmitter $i$ intends to send $d_{ji}$ independent streams to receiver $j$ using a precoding matrix $\mathbf{V}^{[ji]}$ of dimension $A_i\times d_{ji}$. Receiver $j$ zero-forces all the interference  with a receive filter matrix $\mathbf{U}^{[j]}$ of dimension $B_j\times \sum_id_{ji}$. The IA solution requires the simultaneous satisfiability of the following conditions:
\begin{equation}
\label{alignment_con1} \mathbf{U}^{[k]\dag}\mathbf{H}^{[ki]}\mathbf{V}^{[ji]} = \mathbf{0},  \qquad \forall j\neq k
\end{equation}
\begin{equation}
\label{alignment_con1_rank} 
\text{rank}\left(\mathbf{U}^{[j]\dag}\left[\mathbf{H}^{[j1]}\mathbf{V}^{[j1]}, \ldots, \mathbf{H}^{[jM]}\mathbf{V}^{[jM]}\right]\right) = \sum_id_{ji},
~\forall i,j
\end{equation}
where $i\in\{1, 2, \ldots, M\}$, $j, k\in \{1, 2, \ldots, N\}$. $\mathbf{A}^\dag$ denotes the conjugate transpose of the matrix $\mathbf{A}$, and $\left[\mathbf{A}, \mathbf{B}\right]$ represents horizontal concatenation of matrices $\mathbf{A}$ and $\mathbf{B}$.
The IA condition (\ref{alignment_con1_rank}) implies that $\mathbf{V}^{[ji]}$ and $\mathbf{U}^{[j]}$ must have full column rank.

It is well known that linear IA schemes  satisfy reciprocity \cite{Gomadam_Cadambe_Jafar_dist}. In the reciprocal network, the direction of communication is switched, and transmitter $j$ intends to send $\tilde{d}_{ij} = d_{ji}$ DoF to receiver $i$. The precoding filter $\mathbf{\tilde{V}}^{[ij]}$ is obtained by splitting $\mathbf{U}^{[j]}$, i.e., $\mathbf{U}^{[j]} = \left[\mathbf{\tilde{V}}^{[1j]}, \mathbf{\tilde{V}}^{[2j]}, \ldots, \mathbf{\tilde{V}}^{[Mj]}\right]$, where $\mathbf{\tilde{V}}^{[ij]}$ is a $B_j\times d_{ji}$ matrix. The receiving filter $\mathbf{\tilde{U}}^{[i]}$ at receiver $i$ is $\left[\mathbf{V}^{[1i]}, \mathbf{V}^{[2i]}, \ldots, \mathbf{V}^{[Ni]}\right]$ with dimension $A_i\times \sum_{j}d_{ji}$. Then the linear IA feasibility conditions in the reciprocal network are
\begin{equation}
\label{alignment_con1_reciprocity} \mathbf{\tilde{U}}^{[l]\dag}\mathbf{\tilde{H}}^{[lj]}\mathbf{\tilde{V}}^{[ij]} = \mathbf{0}, \qquad \forall i\neq l
\end{equation}
\begin{equation}
\label{alignment_con1_rank_reciprocity} 
\text{rank}\left(\mathbf{\tilde{U}}^{[i]\dag}\left[\mathbf{\tilde{H}}^{[i1]}\mathbf{\tilde{V}}^{[i1]}, \ldots, \mathbf{\tilde{H}}^{[iN]}\mathbf{\tilde{V}}^{[iN]}\right]\right) = \sum_jd_{ji},
~\forall i,j
\end{equation}
where $i, l\in\{1, 2, \ldots, M\}$, $j\in \{1, 2, \ldots, N\}$, and $\mathbf{\tilde{H}}^{[ij]}$ denotes the reciprocal channel from transmitter $j$ to receiver $i$. Also (\ref{alignment_con1_rank_reciprocity}) requires that $\mathbf{\tilde{V}}^{[ij]}$ and $\mathbf{\tilde{U}}^{[i]}$ are full column rank.

{\it Remark:} Note the subtle but essential difference between the MIMO interference channel and MIMO $X$ network, evident in the asymmetric form of the feasibility conditions (\ref{alignment_con1}) (\ref{alignment_con1_rank}) where only one filter matrix ${\bf U}^{[k]}$ is associated with a receiver and accounts for all received messages, but multiple precoding matrices ${\bf V}^{[ji]}$ are associated with each transmitter, one corresponding to each message originating at that transmitter. Similarly, in the reciprocal network, at receiver $i$ multiple precoding matrices $\mathbf{V}^{[ji]}$ in the original network are combined into one receiving filter $\mathbf{\tilde{U}}^{[i]}$, and the receiving filter $\mathbf{U}^{[j]}$ in the original network are split into multiple matrices as the new precoding matrices $\mathbf{\tilde{V}}^{[ij]}$.
This may be also seen as an intriguing form of one-sided decomposition. Note that a two-sided decomposition is also possible, but it would produce a less tight condition, whereas as we will see soon with some examples, the condition obtained with this one-sided decomposition approach will turn out to be tight in the critical test cases (settings with no redundant dimensions).

In the above context, we have the following theorem.
\begin{theorem}
Consider the $M\times N$ user symmetric MIMO $X$ network where each transmitter is equipped with $A$ and each receiver is equipped with $B$ antennas, every transmitter intends to send $d$ DoF to every receiver. If a feasible linear IA solution exists, the symmetric DoF $d$ must satisfy
\begin{equation}
\label{symmetric_case2}
d\leq\frac{A+B}{MN+1}.
\end{equation}
\end{theorem}

{\it Remark:} The result can be shown for arbitrary antenna configurations, without any symmetry assumptions, but is stated here for the symmetric setting where it can be expressed in a compact form. The result essentially states that improper systems are infeasible. The proof is virtually identical to \cite{IC_feasibility_Tse, IC_feasibility}.

Note that since we are only making a claim about ``infeasibility'' we can over-count the variables and still have a valid, albeit less interesting, result. We will, however, check if our threshold is tight through some examples. Note from the recently solved 3 user MIMO interference channel \cite{Wang_Gou_Jafar_3mimoic}, that the settings with no redundancy (where neither $A$ nor $B$ can be reduced without losing DoF) are the ones where the feasibility condition matches the information theoretic DoF value.

{\it Example to Check Tightness of Threshold Value:} Consider a $2\times 2$ user $X$ channel, where each transmitter is equipped with $2$ antennas and each receiver is equipped with $3$ antennas. We already know that $d=1$ is tight here, i.e., information theoretically there is no redundancy in the number of antennas on either side. Also the threshold value from our bound (\ref{symmetric_case2}) is $\frac{2+3}{2\times 2 +1}=1$. This is a good sanity check that our improper condition is not generally loose, i.e., the variables are not being generally over-counted. Similarly, in a $2\times K$ user MIMO $X$ network, where each transmitter is equipped with $K$ antennas and each receiver is equipped with $K+1$ antennas, one can also verify that $d=1$ is tight and there is no redundant antenna dimension. Again our bound (\ref{symmetric_case2})  provides a value $\frac{K+K+1}{2K +1}=1$ which is tight as well.

\section{Proofs: Spatial Scale Invariance and Decomposability of MIMO $X$ Networks}
We only need to prove Theorem $2$ which includes Theorem $1$ as a special case, as stated before. The outer bound proof, presented next, is straightforward and is provided mainly for completeness. The main challenging aspect is the achievability proof presented subsequently.
\subsection{Outer Bound on the DoF of MIMO $X$ networks}
The $M\times N$ user SIMO $X$ network with a single antenna at each transmitter and $R$ antennas at each receiver is considered. The proof for its reciprocal setting, the $N\times M$ MISO $X$ network, follows along the same lines.

\textit{Proof:}
When $M \leq R$, the DoF value for the $X$ network is bounded by the total number of transmit antennas $M$, which is simply the single-user DoF bound.

When $M > R$: If we allow full cooperation among the first $R$ transmitters, then it is equivalent to the $X$ network with $M-R+1$ transmitters, the first transmitter equipped with $R$ antennas and all the other transmitters equipped with single antenna each, and $N$ receivers, each equipped with $R$ antennas. In \cite{Cadambe_Jafar_X}, it is shown that in $X$ network, the number of DoF achieved by all the messages associated with transmitter $m$ or receiver $n$ is upper bounded by $\max(A^t_m,B^r_n)$, where $A^t_m$ and $B^r_n$ stand for the number of antennas at the transmitter $m$ and receiver $n$, respectively. Since allowing cooperation among transmitters does not hurt the capacity, the number of DoF achieved by all the messages associated with the first $R$ transmitters and the receiver $n$ is no more than $\max(R,R)=R$. This gives us the outer bound
\begin{eqnarray}
\sum_{q=1}^{N} \sum_{m=1}^{R} d_{qm} + \displaystyle\sum_{p=R+1}^{M} d_{np} \leq R .
\end{eqnarray}
Repeating the arguments for every $R$ transmitters and each receiver $n$, we arrive at the outer bound of the SIMO $X$ network
\begin{equation}
\sum_{i \in \{1,2 \ldots ,M\}, j \in \{1,2 \ldots ,N\}} d_{ji} \leq \frac{MNR}{M+NR-R}.
\end{equation}
Furthermore, the spatial scaling property of the outer bound is obvious from the derivation presented above.

\subsection{Inner Bound on the DoF of MIMO $X$ networks}\label{sec:inner}
As mentioned previously, the achievability proof of Theorem $2$ is the main challenging aspect. The proof first establishes the achievable DoF for  SIMO and (by reciprocity) MISO $X$ networks, and then uses a one-sided decomposition argument to establish spatial scale invariance for this class of networks.

\textit{Proof:}
When $M \leq R$, beamforming and zero forcing are sufficient to achieve the DoF.

When $M > R$, the achievable scheme is based on interference alignment. Due to the reciprocity of linear beamforming-based alignment, which states that if interference alignment is feasible in the original network then it is also feasible in the reciprocal network, and the achievable DoF are the same between the dual networks \cite{Cadambe_Jafar_X,Gomadam_Cadambe_Jafar_dist}, we only consider the SIMO case. Consider an $n$ symbol extension of the original channel. The value of $n$ will be specified  later. The input-output relationship of the extended channel is described by
\begin{eqnarray}
\mathbf{Y}^{[j]}(\kappa) = \sum_{i=1}^{M} \mathbf{H}^{[ji]}(\kappa) \mathbf{X}^{[i]}(\kappa) +\mathbf{Z}^{[j]}(\kappa)= \sum_{i=1}^{M} \left[\begin{array}{c}\mathbf{H}^{[ji]}_{1}(\kappa)\\
\vdots \\
\mathbf{H}^{[ji]}_{R}(\kappa) \end{array}\right]\mathbf{X}^{[i]}(\kappa)+\mathbf{Z}^{[j]}(\kappa),
~~~j \in \{1,2,\ldots,N\}
\end{eqnarray}
where $\mathbf{X}^{[i]}(\kappa)$ is the $n \times 1$ transmitted signal vector sent from the $i^{th}$ transmitter and $\mathbf{Y}^{[j]}(\kappa)$ is the $nR \times 1$ received signal vector at receiver $j$. $\mathbf{H}^{[ji]}_{r}(\kappa)$ represents the $n \times n$ channel matrix from Transmitter $i$ to the $r^{th}$ receive antenna of Receiver $j$, $r \in \{1,\ldots,R\}$, i.e.,
\begin{eqnarray}
{\mathbf{H}}^{[ji]}_{r}(\kappa)=  \left[ \begin{array}{cccc}  H^{[ji]}_{r}(n(\kappa-1)+1) & 0 & \ldots & 0\\
     0 & H^{[ji]}_{r}(n(\kappa-1)+2) & \ldots & 0\\
    \vdots & \cdots & \ddots & \vdots\\
     0 & 0 & \cdots  & H^{[ji]}_{r}(n\kappa) \\
    \end{array}\right].
\end{eqnarray}
The channel-use index, $\kappa$, is suppressed from now on for compactness. Each transmitter selects the same beamforming matrix $\mathbf{V}_j$ for precoding its symbols intended for Receiver $j$. $\mathbf{V}_j$ is an $n \times |\mathbf{V}_j|$ matrix whose columns are beamforming directions. The number of columns of ${\bf V}_j$, i.e., the value of $|{\bf V}_j|$ will also be specified later in this proof. The transmit signal sent by Transmitter $i$ is $\mathbf{X}^{[i]} = \sum_{j=1}^N \mathbf{V}_j\mathbf{x}^{[ji]}$, where $\mathbf{x}^{[ji]}$ is the $|{\bf V}_j|\times 1$ vector of  $|{\bf V}_j|$  data streams from Transmitter $i$ to Receiver $j$. {\color{black} The received signal at Receiver $l$, wherein $l \in \{1,2,\ldots,N\}$, is expressed as
\begin{eqnarray}
\mathbf{Y}^{[l]} &=& \sum_{i=1}^{M} \left[\begin{array}{c}\mathbf{H}^{[li]}_{1}\\
\vdots \\
\mathbf{H}^{[li]}_{R} \end{array}\right] \left( \sum_{j=1}^N \mathbf{V}_j\mathbf{x}^{[ji]} \right)+\mathbf{Z}^{[l]} \\
&=& \sum_{i=1}^{M}\sum_{j=1}^N \left[\begin{array}{c}\mathbf{H}^{[li]}_{1}\mathbf{V}_j \\
\vdots \\
\mathbf{H}^{[li]}_{R}\mathbf{V}_j \end{array}\right]  \mathbf{x}^{[ji]} +\mathbf{Z}^{[l]} \\
&=& \underbrace{ \sum_{i=1}^{M} \left[\begin{array}{c}\mathbf{H}^{[li]}_{1}\mathbf{V}_l \\
\vdots \\
\mathbf{H}^{[li]}_{R}\mathbf{V}_l \end{array}\right]  \mathbf{x}^{[li]} }_{\mbox{Desired Signal}} + \underbrace{ \sum_{i=1}^{M}\sum_{j=1,j \neq l}^N \left[\begin{array}{c}\mathbf{H}^{[li]}_{1}\mathbf{V}_j  \\
\vdots \\
\mathbf{H}^{[li]}_{R}\mathbf{V}_j \end{array}\right]  \mathbf{x}^{[ji]} }_{\mbox{Interference}} +\mathbf{Z}^{[l]}.
\end{eqnarray}
}

{Consider the symbols desired by Receiver 1, which constitute interference at receivers $l \in \{2,\ldots,N\}$. These symbols are sent by each transmitter along the signal space designated as $\mathbf{V}_1$. Note that because each receiver has $R$ antennas while each transmitter has only 1 antenna, the symbols sent from any $R$ transmitters cannot align among themselves at any receiver. This is because the channel matrix from any $R$ transmitters to the $R$-antenna receiver is invertible almost surely. Therefore, the $\mathbf{V}_1$ spaces from transmitters $1,\ldots,R$, occupy an $R|\mathbf{V}_1|$ dimensional interference space at each undesired receiver $l\in\{2,\ldots, N\}$.  All the remaining undesired $\mathbf{V}_1$ space signals sent from transmitters $R+1,\ldots,M$, are now aligned into these $R|\mathbf{V}_1|$ dimensions as follows.

{\color{black} Let us choose $\mathbf{V}_1 $ to satisfy the following alignment conditions.
\begin{equation}\label{eqn:v1k}
\mathbf{V}_1 \approx \mathbf{H}^{[li]}_{1}\mathbf{V}_{1}  \approx \cdots \approx \mathbf{H}^{[li]}_{R}\mathbf{V}_{1},~~~l \in \{2,\ldots,N\}, i \in \{1,\ldots,M\}.
\end{equation}
Then all the interference due to signals sent along $\mathbf{V}_1$, as seen by receivers $2$ to $N$, will be aligned into the vector space $\underbrace{\mathbf{V}_1 \times \cdots \times \mathbf{V}_1}_\text{$R$ times}$, i.e.,
\begin{equation}
\begin{aligned}
\text{span}\left[ \mathbf{H}^{[li]}\mathbf{V}_{1} \right] =
\text{span}\left[\begin{array}{c}\mathbf{H}^{[li]}_{1}\mathbf{V}_{1} \\ \vdots \\
\mathbf{H}^{[li]}_{R}\mathbf{V}_{1} \end{array}\right] ~{\approx}~
\text{span}\left[\begin{array}{cccc}
\mathbf{V}_{1} & \mathbf{0} & \cdots & \mathbf{0} \\
\mathbf{0} & \mathbf{V}_{1} & \cdots & \mathbf{0} \\
\vdots & \vdots & \ddots & \vdots  \\
\mathbf{0} & \mathbf{0} & \cdots & \mathbf{V}_{1}
\end{array}\right],
\\
 l \in \{2,\ldots,N\}, i \in \{1,\ldots,M\}
\end{aligned}
\end{equation}
}

Similarly, the $N$ messages for Receiver $j$ are sent along the same signal space $\mathbf{V}_j$ by each transmitter and aligned into the same space $\mathbf{V}_j \times \cdots \times \mathbf{V}_j$ at receivers $l \in \{1,\ldots,j-1,j+1,\ldots,N\}$, where they constitute interference. Then we have
\begin{eqnarray}\label{eqn:v2k}
\mathbf{V}_j \approx \mathbf{H}^{[li]}_{r} \mathbf{V}_{j},
~~~ l \in \{1,\ldots,j-1,j+1,\ldots,N\},i \in \{1,\ldots,M\}, r \in \{1,\ldots,R\}.
\end{eqnarray}
{Define $\mathcal{I}_j = \bigcup_{l,i,r} \mathbf{H}^{[li]}_{r}\mathbf{V}_j$, which is the union of all interference terms due to signals transmitted along $\mathbf{V}_j$. The conditions (\ref{eqn:v1k}), (\ref{eqn:v2k}) can now be expressed as $\mathbf{V}_j \approx \mathcal{I}_j$.} These conditions are satisfied simultaneously by the CJ  scheme construction:
\begin{eqnarray}\label{eqn:vk}
\mathbf{V}_j = \bigg\{ \Big(\prod_{l,i,r} (\mathbf{H}^{[li]}_{r})^{\alpha_{r}^{[li]}} \Big) \mathbf{1}, \notag
\text{s. t.} \sum_{l,i,r}\alpha_{r}^{[li]} \leq m, \alpha_{r}^{[li]} \in \mathbb{Z}_+ , \notag \\
 l \in \{1,\ldots,j-1,j+1,\ldots,N\}, i \in \{1,\ldots,M\}, r \in \{1,\ldots,R\}
\bigg\},
\end{eqnarray}
\begin{eqnarray}
\mathcal{I}_j = \bigg\{ \Big(\prod_{l,i,r} (\mathbf{H}^{[li]}_{r})^{\alpha_{r}^{[li]}} \Big) \mathbf{1}, \notag
\text{s. t.} \sum_{l,i,r}\alpha_{r}^{[li]} \leq m+1, \alpha_{r}^{[li]} \in \mathbb{Z}_+ , \notag \\
 l \in \{1,\ldots,j-1,j+1,\ldots,N\}, i \in \{1,\ldots,M\}, r \in \{1,\ldots,R\}
\bigg\}
\end{eqnarray}
where $\mathbf{1}$ is the $n \times 1$ all 1 column vector.
Thus $\mathbf{V}_j$ contains product terms up to degree $m$ and interference $\mathcal{I}_j$ contains product terms up to degree $m+1$. The size of the signal space $\mathbf{V}_j$ (number of column vectors in $\mathbf{V}_j$) and interference $\mathcal{I}_j$, respectively, is
\begin{eqnarray}\label{eqn:v_dim}
|\mathbf{V}_j| = \left( \begin{array}{c} m \\ L \end{array} \right),~|\mathcal{I}_j| = \left( \begin{array}{c} m+1 \\ L \end{array} \right)
\end{eqnarray}
where $L  = MR(N-1)$ is the total number of interference carrying channels. We denote $|\mathbf{V}_j|$ as $|\mathbf{V}|$ and $|\mathcal{I}_j|$ as $|\mathcal{I}|$, because they are the same for all $j$. Notice
\begin{equation}
\frac{|\mathbf{V}|} {|\mathcal{I}|} = \frac{m+1-L}{m+1} \to 1 ~~\text{as} ~~m \to \infty
\end{equation}
 which means $|\mathbf{V}| \approx |\mathcal{I}|$. At Receiver $j$, desired signals occupy $M|\mathbf{V}|$ dimensions and aligned interference occupies $(N-1)R|\mathcal{I}|$ dimensions. To avoid overlaps between desired signals and interference  the size of receive signal space, $nR$, must be at least  as big as the sum of the dimensions of desired signals and interference, $nR \geq M|\mathbf{V}| + R(N-1)|\mathcal{I}|$, so we set $n =  M|\mathbf{V}| / R + (N-1)|\mathcal{I}|$.\footnote{One can guarantee that $n$ is an integer by, e.g., choosing $m=LRz$ wherein $z$ is the    sequence of integers, so that $|{\bf V}|=Rz\binom{LRz-1}{L-1}$ is divisible by $R$.}
 
 Next we  prove the linear independence of the desired signals from interference.  
 
 Let us first simplify the notation as follows.  Relabel all the $L$ interference carrying channels $\mathbf{H}^{[li]}_{r}$ in $\mathbf{V}_j$ as $\mathbf{T}_{1}$ to $\mathbf{T}_{L}$ and their corresponding exponents as $\alpha_1$ to $\alpha_L$. Similar change of notation is also done within all $\mathcal{I}_j$. Then
\begin{equation}
\mathbf{V}_j=\left\{(\mathbf{T}_{1})^{\alpha_1}(\mathbf{T}_{2})^{\alpha_2}\cdots(\mathbf{T}_{L})^{\alpha_L}\mathbf{1}:\sum_{i=1}^L\alpha_{i}\leq m, \alpha_1,\ldots,\alpha_L \in\mathbb{Z}_{+} \right\}.
\end{equation}

Note that $\mathbf{V}_j$ is comprised of column vectors. For ease of exposition, we will  impose a lexicographic order on these columns in the representation of $\mathbf{V}_j$, as follows. First, we arrange all columns from left to right in increasing order of $\alpha_1$. Then for columns of the same $\alpha_1$, we will arrange them in increasing order of $\alpha_2$. In general, given the same tuple $(\alpha_1,\alpha_2,\cdots, \alpha_k)$, $k<L$, we will arrange these columns in increasing order of $\alpha_{k+1}$. For example, consider the setting $L=3$ and $m=5$. Then  $\mathbf{V}_j$ is represented as the matrix
\begin{eqnarray}
\left[\begin{array}{cccccccccc}T_1T_2T_3 &T_1T_2T_3^2& T_1T_2T_3^3& T_1T_2^2T_3&T_1T_2^2T_3^2&T_1T_2^3T_3&T_1^2T_2T_3&T_1^2T_2T^2_3&T^2_1T^2_2T_3&T_1^3T_2T_3\end{array} \right]
\end{eqnarray}
 Such an ordering has the property that a tuple  $(\alpha_1,\alpha_2, \cdots, \alpha_L)$ appears before the tuple $(\beta_1,\beta_2, \cdots, \beta_L)$ if and only if the first $\alpha_i$, which is different from $\beta_i$, is smaller than $\beta_i$.
With this arrangement, we have the following lemma.
\begin{lemma}\label{lemma:VpowerK}
Consider a row vector
\begin{eqnarray}
\mathbf{v}_{j:k}=\left[\begin{array}{cccc}V_1&V_2&\cdots&V_k\end{array}\right]
\end{eqnarray}
which is obtained from the first to the $k^{th}$ column of an arbitrary row of matrix $\mathbf{V}_j$. Now consider a product of the form $\prod_{i=1}^R V_{k_i}$,  $\forall k_i\in\{1,\cdots,k\}$. Note that each product is a monomial in variables of $T_l, l \in \{1,2,\ldots,L\}$. Then, $\prod_{i=1}^R V_{k_i}=(V_k)^R$ if and only if $k_i=k$, for all $i\in\{1,\cdots,R\}$.
\end{lemma}
\proof Suppose $V_k=T_1^{\alpha_1}T_2^{\alpha_2}\cdots T_L^{\alpha_L}$. Then $V_k^R=T_1^{R\alpha_1}T_2^{R\alpha_2}\cdots T_L^{R\alpha_L}$. Suppose $\forall k_i\leq k$ we have $V_{k_i}=T_1^{\beta_1^{[k_i]}}T_2^{\beta_2^{[k_i]}}\cdots T_L^{\beta_L^{[k_i]}}$,  $i\in\{1,\cdots,R\}$  such that $\sum_{i=1}^R\beta_j^{[k_i]}=R\alpha_j, \forall j\in\{1,\cdots, L\}$. According to the ordering of the $V$, since $k_i\leq k$, we have $\beta_1^{[k_i]} \leq \alpha_1$, for all $k_i$. So in order for $\sum_{i=1}^R\beta_1^{[k_1]}=R\alpha_1$, all $\beta_1^{[k_i]}$ have to be equal to $\alpha_1$. Continuing this argument, given $\beta_j^{[k_i]}=\alpha_j$ for all $j<L$, we have $\beta_{j+1}^{[k_i]} \leq \alpha_{j+1}$ for all $k_i$. So in order for $\sum_{i=1}^R\beta_{j+1}^{[k_i]}=R\alpha_{j+1}$,  all $\beta_{j+1}^{[k_i]}$ have to be equal to $\alpha_{j+1}$,  leading to $V_{k_i}=V_k$. \hfill \QED

Without loss of generality, we will prove the linear independence of desired and interfering signal spaces for Receiver 1. Let us define
\begin{eqnarray}
\mathbf{D}^{[1]}_r = \left[\begin{array}{cccc}\mathbf{H}_{r}^{[11]}\mathbf{V}_1 &\mathbf{H}_{r}^{[12]} \mathbf{V}_1 & \cdots & \mathbf{H}_{r}^{[1M]} \mathbf{V}_1\end{array}\right], \quad r \in \{1,\ldots, R\}
\end{eqnarray}
which corresponds to the desired signal at the $r^{th}$ antenna of Receiver 1. Then the desired signal at Receiver 1 is received along the columns of the following matrix,
\begin{eqnarray}
\mathbf{D}^{[1]}=\left[\begin{array}{c}\mathbf{D}_1^{[1]}\\ \mathbf{D}_2^{[1]}\\ \vdots \\ \mathbf{D}_R^{[1]}\end{array}\right].
\end{eqnarray}
Now consider the interference. According to our alignment scheme, the interference signal intended for receiver $l \in \{2,\ldots, N\}$, is aligned into the span of the columns of the following matrix,
\begin{eqnarray}
\mathbf{E}_l =
\left[\begin{array}{cccc}
\mathcal{I}_l & \mathbf{0} & \cdots & \mathbf{0} \\
\mathbf{0} & \mathcal{I}_l & \cdots & \mathbf{0} \\
\vdots & \vdots & \ddots & \vdots  \\
\mathbf{0} & \mathbf{0} & \cdots & \mathcal{I}_l
\end{array}\right]
=\mathbf{I}_R\otimes\mathcal{I}_l, \quad l \in \{2,\ldots, N\}
\end{eqnarray}
where $\mathbf{I}_R$ is the $R\times R$ identity matrix and $\otimes$ denotes the Kronecker product. As a result, all interference signals are aligned into the span of the columns of the following matrix,
\begin{eqnarray}
\mathbf{E}^{[1]}=\left[\begin{array}{ccc}\mathbf{E}_2&\cdots&\mathbf{E}_N\end{array}\right].
\end{eqnarray}
Therefore we need to show the $nR \times nR$ matrix $\mathbf{F}^{[1]} = [\mathbf{D}^{[1]}~~\mathbf{E}^{[1]}]$ has full rank almost surely.
We will show that the desired signals are linearly independent among themselves and the desired signal space does not overlap with the interference space, respectively. 

The difficulty lies in the second step as there is dependency across both columns and rows in the signal space matrix. The columns are dependent because in $X$ networks, desired channels for Receiver $1$ are interfering channels for other receivers. The rows are dependent because we are performing joint MIMO decoding, involving signals that are received at all receive antennas. So we need to perform an induction on both columns and rows at the same time. At each induction step, assuming the original matrix has full rank, we prove the new matrix formed by adding $R$ columns and $R$ rows also has full rank. This is done by identifying a distinct monomial in the polynomial expansion of the determinant. Both the block diagonal structure of the interference and the former lexicographic ordering of the precoding vectors are important in the remainder of the proof, which is described next.

{\em Step 1:} We first prove that the desired signals are linearly independent, i.e., the $nR \times M|\mathbf{V}_1|$ matrix $\mathbf{D}^{[1]}$ has full rank almost surely. To do this, it is sufficient to prove the following $M|\mathbf{V}_1| \times M|\mathbf{V}_1|$ submatrix of $\mathbf{D}^{[1]}$ has full rank almost surely.
\begin{eqnarray}
\mathbf{\bar{D}}^{[1]} = \left[\begin{array}{c}\mathbf{\bar{D}}^{[1]}_1 \\ \mathbf{\bar{D}}^{[1]}_2 \\ \vdots \\ \mathbf{\bar{D}}^{[1]}_R \end{array}\right]
\end{eqnarray}
where
\begin{eqnarray}
\mathbf{\bar{D}}^{[1]}_r = \left[\begin{array}{cccc}\mathbf{\bar{H}}_{r}^{[11]} \mathbf{\bar{V}}_1 &\mathbf{\bar{H}}_{r}^{[12]}\mathbf{\bar{V}}_1 & \cdots & \mathbf{\bar{H}}_{r}^{[1M]}\mathbf{\bar{V}}_1\end{array}\right], \quad r\in\{1,\cdots, R\}
\end{eqnarray}
is comprised of the first $\frac{M|\mathbf{V}_1|}{R}$ rows of $\mathbf{D}^{[1]}_r$, i.e., $\mathbf{\bar{H}}_{r}^{[1i]}$ is a diagonal square matrix of dimension $\frac{M|\mathbf{V}_1|}{R} \times \frac{M|\mathbf{V}_1|}{R}$ obtained from the first $\frac{M|\mathbf{V}_1|}{R}$ rows and columns from matrix $\mathbf{H}_{r}^{[1i]}$ and $\mathbf{\bar{V}}_1$ is the $\frac{M|\mathbf{V}_1|}{R} \times |\mathbf{V}_1|$ matrix obtained from the first $\frac{M|\mathbf{V}_1|}{R}$ rows of matrix $\mathbf{V}_1$. Essentially, we only consider the signals received up to channel use index $\frac{M|\mathbf{V}_1|}{R}$. Note that $\mathbf{\bar{D}}^{[1]}$ has $R$ block rows which correspond to $R$ antennas and $M$ block columns  which correspond to the desired signals from $M$ transmitters.
To prove it is a full rank matrix, we will prove $\det(\mathbf{\bar{D}}^{[1]})\neq 0$ almost surely. The determinant is a polynomial of all channel coefficients up to channel index $\frac{M|\mathbf{V}_1|}{R}$. To prove it is not equal to zero almost surely, it suffices to prove it is not a zero polynomial, which can be proved by showing at least one specific channel realization exists such that the polynomial is not equal to zero. We will set the channel coefficients such that  $\mathbf{\bar{D}}^{[1]}$ becomes a block diagonal matrix with $M$ blocks and each block is a full rank matrix almost surely  which leads to the conclusion that $\mathbf{\bar{D}}^{[1]}$ has full rank almost surely as well. Specifically, consider the $i^{th}$ block column of $\mathbf{\bar{D}}^{[1]}$, i.e.,
\begin{eqnarray}
\left[\begin{array}{c}\mathbf{\bar{H}}_{1}^{[1i]} \mathbf{\bar{V}}_1\\ \mathbf{\bar{H}}_{2}^{[1i]} \mathbf{\bar{V}}_1\\ \vdots \\ \mathbf{\bar{H}}_{R}^{[1i]} \mathbf{\bar{V}}_1 \end{array}\right],\quad i\in\{1,\cdots,M\}
\end{eqnarray}
which corresponds to the desired signal from Transmitter $i$. We set all rows except rows $(i-1)|\mathbf{V}_1|+1, \cdots, i|\mathbf{V}_1|$ of $\mathbf{\bar{D}}^{[1]}$ to zero by setting the corresponding channel coefficients in matrix $\mathbf{\bar{H}}_{r}^{[1i]}$ to zero. This operation involves only channels that originate at Transmitter $i$, so they are independent of other block columns. Note that this can be done because $\mathbf{\bar{V}}_1$ does not contain channel coefficients associated with Receiver 1. As a result, we convert matrix $\mathbf{\bar{D}}^{[1]}$ into a block diagonal matrix where each block is a $|\mathbf{V}_1| \times |\mathbf{V}_1|$ matrix. 

What remains to be shown is that each block is a full rank matrix almost surely. We will prove this by showing that each block matrix satisfies two properties: 1) every entry of each row is a distinct monomial; 2) each row is completely independent of the other rows. If both properties are satisfied, then it follows from Lemma 1 in \cite{Cadambe_Jafar_X} that the matrix has full rank almost surely. It can be easily seen that the first property is satisfied for each row due to the construction of $\mathbf{V}_1$. We only need to prove the second property is satisfied as well. First notice that the rows may not be independent due to the stack of matrix $\mathbf{\bar{V}}_1$ which has $\frac{M|\mathbf{V}_1|}{R}$ rows, corresponding to the signals received at different antennas. In other words, each row of $\mathbf{\bar{V}}_1$ appears periodically with period $\frac{M|\mathbf{V}_1|}{R}$. As a result, if we choose $K$ consecutive rows from $\mathbf{\bar{D}}^{[1]}$, the rows are not independent if and only if $K>\frac{M|\mathbf{V}_1|}{R}$. Now we choose $|\mathbf{V}_1|$ consecutive rows each time and  $|\mathbf{V}_1| < \frac{M|\mathbf{V}_1|}{R}$ because $R<M$. As a result, each row is  independent. Therefore, each block matrix is full rank almost surely. So we have proved the desired signals are linearly independent almost surely.

{\em Step 2:} We will prove that the interference space does not overlap with the signal space. To do that we first reorder the rows and columns of matrix $\mathbf{F}^{[1]}$. The columns of each $\mathbf{E}_l$  are reordered as follows:
\begin{eqnarray}
\mathbf{E}_l=\left[\begin{array}{ccc}\mathbf{I}_R\otimes\mathcal{I}_{l1} &\cdots&\mathbf{I}_R\otimes\mathcal{I}_{l|\mathcal{I}_l|} \end{array}\right], l\in\{2,\cdots,N\}
\end{eqnarray}
where $\mathcal{I}_{lk}$ denotes the $k^{th}$ column of matrix $\mathcal{I}_{l}$. Next, we arrange rows in increasing order of the channel indices. The desired signal received at channel index $\kappa$ is given by
\begin{eqnarray}
\mathbf{D}^{[1]}(\kappa)=\left[\begin{array}{cccc}H_{1}^{[11]}(\kappa) &H_{1}^{[12]}(\kappa) & \cdots & H_{1}^{[1M]}(\kappa)\\
H_{2}^{[11]}(\kappa) &H_{2}^{[12]}(\kappa) & \cdots & H_{2}^{[1M]}(\kappa)\\ \vdots& \vdots &\ddots & \vdots\\
H_{R}^{[11]}(\kappa) &H_{R}^{[12]}(\kappa) & \cdots & H_{R}^{[1M]}(\kappa)
\end{array}\right]\otimes \mathbf{V}_{1}(\kappa)
\end{eqnarray}
where $\mathbf{V}_1(\kappa)$ denotes the $\kappa^{th}$ row of $\mathbf{V}_1$. The interference  caused by messages intended for Receiver $l$ at channel index $\kappa$ is given by the following matrix
\begin{eqnarray}
\mathbf{E}_l(\kappa)=\left[\begin{array}{cccc}\mathcal{I}_{l1}(\kappa)\mathbf{I}_R&\mathcal{I}_{l2}(\kappa) \mathbf{I}_R &\cdots&\mathcal{I}_{l|\mathcal{I}_l|}(\kappa)\mathbf{I}_R \end{array}\right], l\in\{2,\ldots,N\}
\end{eqnarray}
where $\mathcal{I}_{lk}(\kappa)$ denotes the element in the $\kappa^{th}$ row and $k^{th}$ column of matrix $\mathcal{I}_{l}$. As a result, all signals received at channel index $\kappa$ are expressed as
\begin{eqnarray}
\mathbf{F}^{[1]}(\kappa)=\left[\begin{array}{cccc}\mathbf{D}^{[1]}(\kappa)&\mathbf{E}_{2}(\kappa)&\cdots&\mathbf{E}_{N}(\kappa)\end{array}\right].
\end{eqnarray}
After rearranging the rows and columns, the matrix becomes
\begin{eqnarray}
\mathbf{F}^{[1]}=\left[\begin{array}{c}\mathbf{F}^{[1]}(1)\\ \vdots \\ \mathbf{F}^{[1]}(n)\end{array}\right].
\end{eqnarray}

Recall that in Step 1, we already proved that the desired signals are linearly independent almost surely, i.e., the first $M|\mathbf{V}_1|$ columns of $\mathbf{F}^{[1]}$ are linearly independent. This was done by proving the  $M|\mathbf{V}_1| \times M|\mathbf{V}_1|$ matrix $\mathbf{\bar{D}}^{[1]}$ has full rank almost surely. Note that $\mathbf{\bar{D}}^{[1]}$ corresponds to all rows of the desired signals up to channel index $T = \frac{M|\mathbf{V}_1|}{R}$, i.e.,
 \begin{eqnarray}
 \mathbf{\bar{D}}^{[1]} =\left[\begin{array}{c}\mathbf{D}^{[1]}(1)\\ \vdots\\ \mathbf{D}^{[1]}(T)\end{array}\right]
 \end{eqnarray}
after row rearrangements according to channel index.
Next, we will start from $\mathbf{\bar{D}}^{[1]}$, and at each induction step, append $R$ rows and $R$ columns to its bottom and right  in $\mathbf{F}^{[1]}$, and prove that the resulting square matrix has full rank almost surely. These $R$ rows and $R$ columns intersect in an $R \times R$ matrix.
The blocks are added sequentially and
every time the rows in the block correspond to the received signal at channel use index $T+\kappa, \kappa \in\{1,\cdots, (N-1)|\mathcal{I}| \}$. We will arrive at $\mathbf{F}^{[1]}$ in the end.
Now we add the first block, i.e.,
\begin{eqnarray}
\mathbf{G}(1)= \left[\begin{array}{cc}\mathbf{\bar{D}}^{[1]}&\mathbf{B}_{21}\\ \mathbf{D}^{[1]}(T+1)&\mathcal{I}_{21}(T+1)\mathbf{I}_R\end{array}\right]
= \left[\begin{array}{c}\mathbf{D}^{[1]}(1) \\ \vdots\\ \mathbf{D}^{[1]}(T) \\ \mathbf{D}^{[1]}(T+1) \end{array}  \begin{array}{c}\mathcal{I}_{21}(1)\mathbf{I}_R\\ \vdots \\ \mathcal{I}_{21}(T) \mathbf{I}_R  \\ \mathcal{I}_{21}(T+1) \mathbf{I}_R \end{array} \right].
\end{eqnarray}
We will now prove that $\mathbf{G}(1)$ has full rank, i.e., $\det(\mathbf{G}(1))\neq 0$ almost surely. The entries in $\mathbf{D}^{[1]}(T+1)$ and $\mathcal{I}_{21}(T+1)\mathbf{I}_R$ are independent of the entries in $\mathbf{\bar{D}}^{[1]}$ and $\mathbf{B}_{21}$ because they correspond to different channel uses. Fix $\mathbf{\bar{D}}^{[1]}$ and $\mathbf{B}_{21}$, then $\det(\mathbf{G}(1))$ is a polynomial in variables of the entries in $\mathbf{D}^{[1]}(T+1)$ and $\mathcal{I}_{21}(T+1)\mathbf{I}_R$. Each term in polynomial $\det(\mathbf{G}(1))$ is a product of $R$ entries, each chosen from a distinct row and distinct column of $\left[\mathbf{D}^{[1]}(T+1)~\mathcal{I}_{21}(T+1)\mathbf{I}_R\right]$. One of these is the term $\det(\mathbf{\bar{D}}^{[1]})$$(\mathcal{I}_{21}(T+1))^R$. To prove $\det(\mathbf{G}(1))\neq0$ almost surely, it is sufficient to prove it is not a zero polynomial, which can be proved if $(\mathcal{I}_{21}(T+1))^R$ is a unique monomial. Since $\mathbf{D}^{[1]}(T+1)$ contains channel coefficients associated with Receiver 2 while $(\mathcal{I}_{21}(T+1))^R$ does not contain those coefficients, we have to choose all $R$ entries from $\mathcal{I}_{21}(T+1)\mathbf{I}_R$ to produce $(\mathcal{I}_{21}(T+1))^R$. Therefore, it is a unique monomial and $\mathbf{G}(1)$ has full rank.

We proceed similarly to add the $\kappa^{th}$ block, $\kappa\in\{2,\cdots, (N-1)|\mathcal{I}| \}$, i.e., 
\begin{eqnarray}\label{eq:iterativeH}
\mathbf{G}(\kappa)=\left[\begin{array}{cc}\mathbf{G}(\kappa-1)&\mathbf{B}(\kappa-1)\\ \mathbf{C}(\kappa)&\mathcal{I}_{lk}(T+\kappa) \mathbf{I}_R\end{array}\right].
\end{eqnarray}
where $l=( \lceil\frac{\kappa}{|\mathcal{I}|}\rceil+1 )$, $k=(\kappa - (l-2)|\mathcal{I}|)$
and
\begin{eqnarray}
\mathbf{C}(\kappa) =\left[\begin{array}{ccccc}\mathbf{D}^{[1]}(T+\kappa)& \mathbf{E}_2(T+\kappa)&\cdots&\mathbf{E}_{l-1}(T+\kappa) \end{array} \begin{array}{cccccc}\mathcal{I}_{l1}(T+\kappa)\mathbf{I}_R&\cdots&\mathcal{I}_{l(k-1)}(T+\kappa)\mathbf{I}_R\end{array} \right], 
\\
\mathbf{B}(\kappa-1) =\left[\begin{array}{ccc}\mathcal{I}_{lk}(1) & \cdots & \mathcal{I}_{lk}(T+\kappa-1) \end{array}\right]^T.
\end{eqnarray}
Next, we will use induction to prove $\mathbf{G}(\kappa)$ is full rank almost surely.  Assuming $\mathbf{G}(\kappa-1)$ is full rank almost surely, we will prove $\det(\mathbf{G}(\kappa))\neq 0$ almost surely. 
Notice that $[\mathbf{C}(\kappa)~ \mathcal{I}_{lk}(T+\kappa) \mathbf{I}_R]$ is independent of $[\mathbf{G}(\kappa-1)~\mathbf{B}(\kappa-1)]$.  Fix $[\mathbf{G}(\kappa-1)~\mathbf{B}(\kappa-1)]$, now the determinant becomes a polynomial in variables of $\mathbf{C}(\kappa)$ and $\mathcal{I}_{lk}(T+\kappa) \mathbf{I}_R$. It is sufficient to prove it is not a zero polynomial. Each term in the polynomial is a product of $R$ entries, each chosen from one different row and one different column of $[\mathbf{C}(\kappa)~\mathcal{I}_{lk}(T+\kappa) \mathbf{I}_R]$. And the polynomial contains the term $\det(\mathbf{G}(\kappa-1))(\mathcal{I}_{lk}(T+\kappa))^R$. If we can prove that $(\mathcal{I}_{lk}(T+\kappa))^R$ is a unique monomial, then the polynomial is not a zero polynomial since $\det(\mathbf{G}(\kappa-1)) \neq 0$, by induction assumption. We will now prove that indeed $(\mathcal{I}_{lk}(T+\kappa))^R$ is a unique monomial. Note that $(\mathcal{I}_{lk}(T+\kappa))^R$ does not contain channel coefficients associated with Receiver $l$ while all entries in $[\begin{array}{ccccc}\mathbf{D}^{[1]}(T+\kappa)& \mathbf{E}_2(T+\kappa)&\cdots&\mathbf{E}_{l-1}(T+\kappa)\end{array}]$ contain those coefficients. Therefore, in order to make the product to be the same as $(\mathcal{I}_{lk}(T+\kappa))^R$, columns of $[\begin{array}{ccccc}\mathbf{D}^{[1]}(T+\kappa)& \mathbf{E}_2(T+\kappa)&\cdots&\mathbf{E}_{l-1}(T+\kappa)\end{array}]$ cannot be chosen. As a result, we only consider choosing $R$ entries from different columns and rows of $\left[\begin{array}{cccccc}\mathcal{I}_{l1}(T+\kappa)\mathbf{I}_R&\cdots&\mathcal{I}_{lk}(T+\kappa)\mathbf{I}_R\end{array}\right]$.
Essentially, the problem becomes to pick $R$ entries each arbitrarily from the vector $[ \mathcal{I}_{l1}(T+\kappa)~ \mathcal{I}_{l2}(T+\kappa)\cdots  \mathcal{I}_{lk}(T+\kappa)]$ and prove the product is equal to $( \mathcal{I}_{lk}(T+\kappa))^R$ if and only if $\mathcal{I}_{lk}(T+\kappa)$ is chosen every time. Mathematically, we want to prove that $\prod_{i=1}^{R} \mathcal{I}_{lk_i}(T+\kappa)$, $k_i\in\{1,\cdots, k\}$ is equal to $( \mathcal{I}_{lk}(T+\kappa))^R$ if and only if $k_i=k$, for all $i\in\{1,\cdots,R\}$. From Lemma \ref{lemma:VpowerK}, this is indeed true. Therefore, we arrive at the conclusion that $( \mathcal{I}_{lk}(T+\kappa))^R$ is a unique monomial and $\mathbf{G}(\kappa)$ has full rank almost surely. Following the induction on $\kappa$ up to its final value, $\kappa=(N-1)|\mathcal{I}|$, we have $\mathbf{G}(\kappa) = \mathbf{F}^{[1]}$. Thus, we conclude that the interference space does not overlap with the signal space.

Therefore, the  accessible DoF for each receiver  equal $R\frac{M|\mathbf{V}|}{nR} = R\frac{M|\mathbf{V}|}{M|\mathbf{V}| + R(N-1)|\mathcal{I}|} \to \frac{MR}{M+NR-R}$ as $m \to \infty$, resulting in a sum DoF of $\frac{MNR}{M+NR-R}$, as desired. At this point we have completed the proof of our DoF result for SIMO and MISO $X$ networks.


We now prove the spatial scale invariance property for extensions of MISO or SIMO $X$ networks.  Let us scale the number of antennas at each node by a factor of $A$  and prove the DoF also scale  by a factor of $A$. When $M \leq R$, the achievable scheme involves only zero forcing and it is easy to see that the DoF scale with $A$.
When $M>R$, {\color{black} we establish spatial scale invariance for the SIMO $X$ network by a decomposition argument and the MISO case follows by reciprocity.} For the spatially scaled SIMO $X$ network, we use transmitter side decomposition. Transmitter side decomposition means that we view each transmitter with $A$ antennas as $A$ distributed transmitters with a single antenna each, such that each of these $A$ transmitters has an independent message for each of the $N$ receivers. In other words, we do not allow joint processing of signals among the $A$ antennas at each transmitter. Then we obtain an $AM \times N$ user SIMO $X$ network with a single antenna at each transmitter and $AR$ antennas at each receiver rather than an $M \times N$ user MIMO $X$ network with $A$ antennas at each transmitter and $AR$ antennas at each receiver. By the result established for  SIMO $X$ networks, $\frac{AMNAR}{AM+NAR-AR}=A\left(\frac{MNR}{M+NR-R}\right)$ DoF are  achieved almost surely. This completes the proof. 


\section{Proofs: Linear Feasibility of MIMO $X$ Networks}
In Section \ref{linear_feasibility}, it is shown that to achieve linear IA in MIMO $X$ network, (\ref{alignment_con1}) and (\ref{alignment_con1_rank}) should be satisfied simultaneously. In our channel model, the MIMO channels are generic and hence have no structure. Therefore, condition (\ref{alignment_con1_rank}) implies that the precoding and receiving filters must have full column rank. Then due to the duality between the original channel and the reciprocal channel, in the following proof of Theorem 3, we only consider condition (\ref{alignment_con1}), i.e.,  $\mathbf{U}^{[k]\dag}\mathbf{H}^{[ki]}\mathbf{V}^{[ji]} = \mathbf{0}, \forall i\in\{1, 2, \ldots, M\},j\in\{1, 2, \ldots, N\},k\in\{1, \ldots, j-1, j+1, \ldots, N\}$.

\textit{Proof:}
The proof of Theorem 3 consists of two steps. First we derive the properness conditions for the symmetric DoF $d$ in MIMO $X$ network, then we prove that in MIMO $X$ network, improper implies infeasible.

Similar to interference channel \cite{IC_feasibility_TSP}, we can obtain the total number of scalar equations in (\ref{alignment_con1}) as
\begin{equation}
\label{N_e}
N_e=\sum_{k=1}^{N}[(\sum_{i=1}^{M}d_{ki})(\sum_{j=1,j\neq k}^{N}\sum_{i=1}^{M}d_{ji})]
\end{equation}

In the first step, when counting the variables in (\ref{alignment_con1}), we need to remove the superfluous variables that do not help with IA. At the receiver, according to \cite{IC_feasibility_TSP}, for the matrix $\mathbf{U}^{[k]}$, we can find one invertible matrix $\mathbf{P}^{[k]}$ with dimension $\sum_id_{ki}\times \sum_id_{ki}$ satisfying
\begin{equation}
\label{U_transform} \mathbf{U}^{[k]}{\mathbf{P}^{[k]}}^{-1} = \left[
                                 \begin{array}{c}
                                   \mathbf{I} \\
                                   \mathbf{\hat{U}}^{[k]} \\
                                 \end{array}
                               \right]
\end{equation}
where $\mathbf{\hat{U}}^{[k]}$ is a $(B_k-\sum_id_{ki})\times \sum_id_{ki}$ matrix. It is easy to argue that the linearly independent columns of $\mathbf{U}^{[k]}$ and those of $\mathbf{U}^{[k]}{\mathbf{P}^{[k]}}^{-1}$ span the same space, and the latter is the basis with the fewest variables for such space. Similarly, for the matrix $\mathbf{V}^{[ji]}$, we can find one invertible matrix $\mathbf{Q}^{[ji]}$ of dimension $d_{ji}\times d_{ji}$
\begin{equation}
\label{V_transform} \mathbf{V}^{[ji]}{\mathbf{Q}^{[ji]}}^{-1} = \left[
                                 \begin{array}{c}
                                   \mathbf{I} \\
                                   \mathbf{\hat{V}}^{[ji]} \\
                                 \end{array}
                               \right]
\end{equation}
where $\mathbf{\hat{V}}^{[ji]}$ is a $(A_i-d_{ji})\times d_{ji}$ matrix. Therefore, after removing the superfluous variables, the total number of variables is
\begin{equation}
\label{N_v}
N_v=\sum_{j=1}^{N}\sum_{i=1}^{M}(A_i-d_{ji})d_{ji}+\sum_{j=1}^{N}[(B_j-\sum_{i=1}^{M}d_{ji})\sum_{i=1}^{M}d_{ji}].
\end{equation}

Then we count the number of equations in the the IA condition (\ref{alignment_con1}), which can be rewritten as
\begin{equation}
\label{new_cond}
\mathbf{U}^{[k]\dag}\mathbf{H}^{[ki]}\mathbf{\overline{V}}^{[i]} = \mathbf{0},
\end{equation}
where $\mathbf{\overline{V}}^{[i]} = [\mathbf{V}^{[1i]},\mathbf{V}^{[2i]},...,\mathbf{V}^{[k-1,i]},\mathbf{V}^{[k+1,i]},...,\mathbf{V}^{[Ni]}]$. For $\mathbf{\overline{V}}^{[i]}$, we can also find one invertible matrix $\mathbf{\overline{Q}}^{[i]}$ of dimension $\sum_{j\neq k}d_{ji}\times \sum_{j\neq k}d_{ji}$
\begin{equation}
\label{new_V_transform} \mathbf{\overline{V}}^{[i]}{\mathbf{\overline{Q}}^{[i]}}^{-1} = \left[
                                 \begin{array}{c}
                                   \mathbf{I} \\
                                   \mathbf{\widetilde{V}}^{[i]} \\
                                 \end{array}
                               \right]
\end{equation}
where $\mathbf{\widetilde{V}}^{[i]}$ is a $(A_i-\sum_{j\neq k}d_{ji})\times \sum_{j\neq k}d_{ji}$ matrix. Obviously, each entry in $\mathbf{\widetilde{V}}^{[i]}$ can be expressed as a function of entries in $\mathbf{\hat{V}}^{[ji]}$. For the channel matrix, it can be partitioned as
\begin{equation}
\label{channel}
\mathbf{H}^{[ki]} = \left[
                            \begin{array}{cc}
                            \mathbf{H}^{[ki]}_{(1)} & \mathbf{H}^{[ki]}_{(2)} \\
                            \mathbf{H}^{ki}_{(3)} &  \mathbf{H}^{[ki]}_{(4)} \\
                            \end{array}
                          \right]
\end{equation}
where $\mathbf{H}^{[ki]}_{(1)}$ is a $\sum_id_{ki}\times \sum_{j\neq k}d_{ji}$ matrix. Now the linear IA condition in (\ref{new_cond}) can be expressed as
\begin{equation}
\label{alignment_con2}
{\mathbf{P}^{[k]\dag}}
                          \left[
                            \begin{array}{cc}
                              \mathbf{I} & {\mathbf{\hat{U}}^{[k]\dag}} \\
                            \end{array}
                          \right]
                          \left[
                            \begin{array}{cc}
                            \mathbf{H}^{[ki]}_{(1)} & \mathbf{H}^{[ki]}_{(2)} \\
                            \mathbf{H}^{[ki]}_{(3)} &  \mathbf{H}^{[ki]}_{(4)} \\
                            \end{array}
                          \right]
                          \left[
                            \begin{array}{c}
                            \mathbf{I }\\
                            \mathbf{\widetilde{V}}^{[i]} \\
                            \end{array}
                          \right]
                          \mathbf{\overline{Q}}^{[i]} = 0.
\end{equation}
Since $\mathbf{P}^{[k]}$ and $\mathbf{\overline{Q}}^{[i]}$ are both invertible, we can get
\begin{equation}
\label{align_con3}
\mathbf{H}^{[ki]}_{(1)} + {\mathbf{\hat{U}}^{[k]\dag}}\mathbf{H}^{[ki]}_{(3)}+\mathbf{H}^{[ki]}_{(2)}\mathbf{\widetilde{V}}^{[i]}+{\mathbf{\hat{U}}^{[k]\dag}}\mathbf{H}^{[ki]}_{(4)}\mathbf{\widetilde{V}}^{[i]}=0.
\end{equation}
It's easy to verify that the total number of scalar equations in (\ref{align_con3}) is
\begin{equation}
\label{N_e}
N_e=\sum_{k=1}^{N}[\sum_{i=1}^{M}d_{ki}(\sum_{j=1,j\neq k}^{N}\sum_{i=1}^{M}d_{ji})].
\end{equation}
In the symmetric system described in Section 4, according to (\ref{N_v}) and (\ref{N_e}), the total number of equations and variables are
\begin{equation}
\label{symmetric2_Nv}
N_v=MNd[A+B-(M+1)d]
\end{equation}
\begin{equation}
\label{symmetric2_Ne}
N_e=M^2Nd^2(N-1).
\end{equation}
If the system is proper, i.e., $N_e \leq N_v$, (\ref{symmetric_case2}) must be satisfied.

The final step, proving that in MIMO $X$ network improper systems are infeasible, uses transcendental field extension theory and mirrors the proof presented in \cite{IC_feasibility}. Here we assume $N_e > N_v$ and the equivalent interference alignment condition in (\ref{align_con3}) is satisfied. If then we get a contradiction, the proof is completed, which means that when (\ref{align_con3}) is satisfied, $N_e$ cannot be larger than $N_v$.

We consider the filed $F$ defined over complex numbers $\mathbb{C}$, consisting of all the rational functions of entries in the matrix $\mathbf{\hat{U}}^{[k]}$ ($k\in\{1,2,\ldots,N\}$) and $\mathbf{\hat{V}}^{[ji]}$ ($i\in\{1,2,\ldots,M\}$, $j\in\{1,2,\ldots,N\}$). Therefore, the entries in all the matrices $\mathbf{\hat{U}}^{[k]}$ and $\mathbf{\hat{V}}^{[ji]}$ form the transcendence basis of $F$, and the transcendence degree of $F$ is equal to the number of entries in all matrices $\mathbf{\hat{U}}^{[k]}$ and $\mathbf{\hat{V}}^{[ji]}$, i.e., $N_v$. Then we define the matrix $\mathbf{F}_{k,i}(\mathbf{\hat{U}},\mathbf{\hat{V}})$ as follows
\begin{equation}
\label{F}
\mathbf{F}_{k,i}(\mathbf{\hat{U}},\mathbf{\hat{V}}) = -[\mathbf{H}^{[ki]}_{(2)}\mathbf{\widetilde{V}}^{[i]}+{\mathbf{\hat{U}}^{[k]\dag}}\mathbf{H}^{[ki]}_{(3)} + {\mathbf{\hat{U}}^{[k]\dag}}\mathbf{H}^{[ki]}_{(4)}\mathbf{\widetilde{V}}^{[i]}].
\end{equation}
Note that in $\mathbf{F}_{k,i}(\mathbf{\hat{U}},\mathbf{\hat{V}})$, each entry is a quadratic polynomial function of entries in the matrices $\mathbf{\hat{U}}^{[k]}$ and $\mathbf{\hat{V}}^{[ji]}$. Therefore, the entries in $\mathbf{F}_{k,i}(\mathbf{\hat{U}},\mathbf{\hat{V}})$ belong to the field $F$. We can find the number of quadratic polynomials in $\mathbf{F}_{k,i}(\mathbf{\hat{U}},\mathbf{\hat{V}})$ is equal to $N_e$. When $N_e > N_v$, the quadratic polynomials in $\mathbf{F}_{k,i}(\mathbf{\hat{U}},\mathbf{\hat{V}})$ is algebraically dependent over $F$. Then we can find a nonzero polynomial $p$ satisfying
\begin{equation}
\label{p}
p(\mathbf{F}_{1,1}(\mathbf{\hat{U}},\mathbf{\hat{V}}),\mathbf{F}_{1,2}(\mathbf{\hat{U}},\mathbf{\hat{V}}),...,\mathbf{F}_{N,M}(\mathbf{\hat{U}},\mathbf{\hat{V}})) = 0
\end{equation}
for all $\mathbf{\hat{U}}^{[k]}$ and $\mathbf{\hat{V}}^{[ji]}$. Here, it is worthwhile noticing that the polynomial $p$ is only dependent on the matrices $\mathbf{H}^{[ki]}_{(2)}$, $\mathbf{H}^{[ki]}_{(3)}$ and $\mathbf{H}^{[ki]}_{(4)}$, but independent of $\mathbf{H}^{[ki]}_{(1)}$. For the polynomial $p$, when viewed as a polynomial of variables $(\mathbf{H}^{[11]}_{(1)},\mathbf{H}^{[12]}_{(1)},...,\mathbf{H}^{[NM]}_{(1)})$, we can do the local expansion at $(\mathbf{F}_{1,1}(\mathbf{\hat{U}},\mathbf{\hat{V}}),\mathbf{F}_{1,2}(\mathbf{\hat{U}},\mathbf{\hat{V}}),...,\mathbf{F}_{N,M}(\mathbf{\hat{U}},\mathbf{\hat{V}}))$,

\begin{equation}
\label{local_expansion}
\begin{aligned}
&p(\mathbf{H}^{[11]}_{(1)},\mathbf{H}^{[12]}_{(1)},...,\mathbf{H}^{[NM]}_{(1)})\\=
&p(\mathbf{F}_{1,1}(\mathbf{\hat{U}},\mathbf{\hat{V}}),\mathbf{F}_{1,2}(\mathbf{\hat{U}},\mathbf{\hat{V}}),...,\mathbf{F}_{N,M}(\mathbf{\hat{U}},\mathbf{\hat{V}})) +\sum Tr(\mathbf{H}^{[ki]}_{(1)}-\mathbf{F}_{k,i}(\mathbf{\hat{U}},\mathbf{\hat{V}}))\mathbf{R}_{k,i}
\end{aligned}
\end{equation}
where $Tr$ denotes the trace operation and $\mathbf{R}_{k,i}$ is a matrix whose entries are polynomials dependent on the entries of $\mathbf{F}_{k,i}(\mathbf{\hat{U}},\mathbf{\hat{V}})$ and the coefficients of $p$ only. Due to (\ref{p}), we have
\begin{equation}
\label{local_expansion_conti}
p(\mathbf{H}^{[11]}_{(1)},\mathbf{H}^{[12]}_{(1)},...,\mathbf{H}^{[NM]}_{(1)})=
\sum Tr(\mathbf{H}^{[ki]}_{(1)}-\mathbf{F}_{k,i}(\mathbf{\hat{U}},\mathbf{\hat{V}}))\mathbf{R}_{k,i}.
\end{equation}
According to (\ref{align_con3}) and (\ref{F}), finally we can obtain
\begin{equation}
\label{p_contraction}
p(\mathbf{H}^{[11]}_{(1)},\mathbf{H}^{[12]}_{(1)},...,\mathbf{H}^{[NM]}_{(1)}) = 0.
\end{equation}
Recall that the polynomial $p$ is independent of the matrices $\mathbf{H}^{[ki]}_{(1)}$. In our channel model, $\mathbf{H}^{[ki]}_{(1)}$ is generic. It implies that the equation (\ref{p_contraction}) cannot hold unless $p$ is equal to zero identically, which contradicts that $p$ is a non-zero polynomial. This completes the proof.


\section{Conclusion}
 In this work, we close the open problem of finding the DoF of MIMO $X$ networks with $A$ antennas at each node. In particular we settle the spatial scale invariance conjecture for this class of networks, as well as SIMO and MISO $X$ networks. In terms of the achievable scheme, we reveal a one-sided decomposability property of  $X$ networks. As a minor addendum, we  explore the feasibility of linear interference alignment based only on spatial beamforming, and prove that improper systems are infeasible, by extending previous work on interference networks.

\end{document}